\documentstyle[prl,aps,twocolumn,epsf]{revtex}
\begin{document}
\twocolumn[\hsize\textwidth\columnwidth\hsize\csname@twocolumnfalse%
\endcsname
\title{Hyperelliptic curves for multi-channel quantum wires\\
and the multi-channel Kondo problem}
\author{P. Fendley$^1$ and H. Saleur$^2$} 
\address{$^1$ Physics Department, University of
Virginia, Charlottesville VA 22901}
\address{$^2$ Department of
Physics, University of Southern California, Los Angeles CA 90089-0484}
\maketitle
\begin{abstract}

We study the current in a multi-channel quantum wire and the
magnetization in the multi-channel Kondo problem. We show that at zero
temperature they can be written simply in terms of contour integrals
over a (two-dimensional) hyperelliptic curve. This allows one to
easily demonstrate the existence of weak-coupling to strong-coupling
dualities. In the Kondo problem, the curve is the same for under- and
over-screened cases; the only change is in the contour.

\end{abstract}
\pacs{PACS numbers: ???} ]

\section{Introduction}

In view of the importance of understanding non-fermi-liquid behavior,
a wide variety of different one-dimensional models have been
studied. Two of the most interesting are the Luttinger liquid and the
multi-channel Kondo problem, where a variety of non-perturbative 
properties have been observed, both theoretically and experimentally.

The methods to study such systems are few. Conformal invariance can be
used to understand fixed points and their neighborhoods. In the
integrable cases (that turn out to be far more common than could have
been hoped for), the Bethe ansatz in principle gives access to all
quantities of interest, including the cross-over behavior of Green
functions. In practice, many computations are remarkably or impossibly
difficult to carry out (see \cite{AFL,Korepin}; \cite{FLSbig,LSS} for
more recent progress), and more direct methods are definitely desired.

Duality could be such a method. In the different context of
supersymmetric gauge theories, the last couple of years have witnessed
astonishing progress on non-perturbative questions, following the
seminal work of \cite{SW} where a certain form of duality (combined
with analyticity) was first exploited.  It seems reasonable to hope
that some of these ideas could be used in the context of condensed
matter and statistical physics, maybe providing a new, more elegant
way of using the Bethe ansatz, and maybe allowing one to solve, at
least partially, more general classes of problems.

Some progress in this direction has been accomplished in
\cite{Paul,FS,FSi}, where it was shown that various properties of the
Kondo problem at arbitrary spin and the Luttinger tunneling problem
did exhibit remarkable representations in terms of hyperelliptic
curves. These representations give rise to various forms of duality,
and to direct reformulations of the Bethe ansatz in terms of monodromy
and differential equations.

Our goal in this note is to extend some of these results to the
multichannel case. We will show in particular that an exact duality
relation for the current holds in the multichannel quantum wire case,
generalizing results of \cite{FLSbig}, and discuss how Fermi and
non-Fermi liquid Kondo fixed points have a unifying representation in
terms of contour integrals over hyperelliptic curves.

\section{The multi-channel quantum wire}

We first consider a problem with $k$ species of electrons (flavors) in
one dimension, with a charge interaction and a single impurity. In
general, we refer to this problem as a ``multichannel quantum wire'',
though the model might need to be refined to describe experimental
situations for general values of $k$.  The case $k$=1 corresponds to
edge states in the fractional quantum Hall effect \cite{Moon}, the
case $k$=2 to quantum wires (with spin isotropy) \cite{KF}, the case
$k$=4 presumably to armchair nanotubes \cite{nano}.

Without impurity, the action is made up of a free-fermion part and 
a charge interaction
\begin{equation}
H=\pi\int dx \left[J_L^2+J_R^2+g_{Lutt}J_LJ_R\right]\label{start}
\end{equation}
where the charge-density current for the left movers is
$J_L=\sum_{i=1}^k \psi_{iL}^\dagger\psi_{iL}$,
and likewise for right movers.  Coupling an impurity to the electrons
adds a scattering term
\begin{equation}
\delta H=\int dx~ V(x)\sum_{i=1}^k\psi_i^\dagger\psi_i(x=0),
\label{delH}
\end{equation}
where the potential $V$ takes negligible values away from the origin.
Like the cases $k=1$ or $k=2$ treated in detail elsewhere
\cite{FLSbig,LeSSii}, we can bosonize and perform the usual
decomposition into odd and even fields. This yields a theory defined
on the full line with a purely chiral interaction at the origin. Calling
$\Theta_i$ the boson associated with the original fermion $\psi_i$, we
introduce new fields
\begin{eqnarray}
\Phi&=&{i\over k}\sum_{i=1}^k\Theta_i\nonumber\\
\Phi_j&=&\Theta_j-{1\over k}\sum_{i=1}^k\Theta_i.
\end{eqnarray}
In the bosonic formulation, the fermionic-interaction parameter
$g_{Lutt}$ in (\ref{start}) is replaced by the usual Luttinger
parameter $g$, defined here so that $g=k$ corresponds to the
non-interacting case $g_{Lutt}=0$. The complete action now reads
\begin{eqnarray}
S=&&{1\over 16\pi}\int dx \sum_{j=1}^k \left[
\left(\partial_x\Phi_j\right)^2+ \left(\Pi_j\right)^2\right]\nonumber\\
&+&{k^2\over 16\pi g}\int dx  \left[
\left(\partial_x\Phi\right)^2+ \left(\Pi\right)^2\right]\nonumber\\ 
&+& \lambda \left(e^{i\phi(0)}\sum_{j=1}^k e^{i\phi_j(0)}+cc\right)
\label{fullint}
\end{eqnarray}
where $\phi$ denotes the right-moving component of $\Phi$,
and $\sum_{j=1}^k \phi_j=0$.
The chiral propagators  are
\begin{eqnarray}
\nonumber\langle \phi(z)\phi(w)\rangle&=&-{2g\over k^2}\ln(z-w)\\
\nonumber\langle\phi_i(z)\phi_i(w)\rangle&=&-{k-1\over k}\ln(z-w)\\
\nonumber\langle\phi_i(z)\phi_j(w)\rangle&=&{1\over k}~\ln(z-w), i\neq j\\
\nonumber\langle\phi_i(z)\phi(w)\rangle&=&0
\end{eqnarray}

In the last sum in (\ref{fullint}), we recognize the well-known
bosonic expression of the fundamental parafermions $\chi_1$ in the $Z_k$ theory
\cite{FZ,GN}, so the interaction term can be written equivalently as
\begin{equation}
\lambda\left(e^{i\phi(0)}\chi_1(0)+
e^{-i\phi(0)}\chi_1^\dagger(0)\right)\label{paraf}.
\end{equation}
One can fold back this theory to obtain what one might call the
``level $k$'' generalization of the boundary sine-Gordon model -- that
is, the boundary version of the generalized supersymmetric, or level
$k$ sine-Gordon model well studied in the literature in the bulk case
\cite{BL}. The case $k=2$ is the ordinary $N$=1 supersymmetric
sine-Gordon model, and additional details of these various
manipulations can be found in \cite{LeSSii}. The dimension of the
perturbing operator is $d=1-{1\over k}+{g\over k^2}$; it is marginal
at the non-interacting point $g=k$, and relevant (irrelevant) for
$g<k$ ($g>k$).

An interesting way to write the interaction (\ref{paraf}) is in the
form $J^++J^-$, where the $J$ are deformed $SU(2)_k$ currents; this
exhibits a relation with the $k$-channel Kondo model to be discussed
in the next section. This is completely analogous to the $k=1$ case
studied in detail in \cite{FLS,BLZ}, where the $J^\pm$ are the usual
vertex operators.

It is worthwhile to comment briefly on cocycles (Klein factors)
here, which have to be handled with great care in problems
with several fermion species. In addition to the usual exponential 
of a free boson, each fermion $\psi_{i,LR}$ requires a real cocycle
$\eta_{i,LR}$ such that $\{\eta_{i,C},\eta_{j,C'}\}=\delta_{ij}\delta_{CC'}$.
In the last term in (\ref{fullint}), this means that
each exponential $e^{i\phi_j(0)}$ should come up, in fact, with 
a prefactor $\eta_{jL}\eta_{jR}$. However, because of charge neutrality,
and the fact that pairs of fermions commute, these factors disappear from
the computation of physical quantities like the free energy
or the conductance, and can be safely ignored.   

The method of \cite{FLSbig} and \cite{LeSSii} can be easily
generalized to compute the current using Bethe ansatz and massless
scattering \footnote{There are subtle issues about charging effects or
 the way the voltage
difference would be imposed in a real quantum wire, that we
intend to discuss elsewhere. In the present paper, $V$ simply controls
the difference of populations of left and right movers in formal analogy with 
the $k=1$ Hall case.}. The current at $T=0$ can be found explicitly by
using the Wiener-Hopf method. The formulas of \cite{FLSbig} and
\cite{LeSSii} generalize to the case of $k$ channels, with a $k$
dependent kernel (using the notations of \cite{LeSSii})
$$
N(\omega)=\sqrt{2\pi({1\over k}+h')} {\Gamma[i(1+kh')\omega/2h']
\Gamma(i\omega/2) \over \Gamma (i\omega/2h')
\Gamma(ik\omega/2)
\Gamma\left({1\over 2}+{i\omega\over 2}\right)}e^{i\omega\Lambda}
$$
where to parameterize the interactions we introduce
$$h'={1\over g}-{1\over k}$$ 
($h'$ is denoted $1/\gamma$ in
\cite{LeSSii}). The parameter $g$ is the conductance without impurity
(in units of $e^2/h$), so $g=k$ for $k$ channels of free electrons.

The interactions are parameterized by
$u\propto V\lambda^{1/h}$. 
where we have defined the coupling $h$ ``dual'' to $h'$ by
$$h={g\over k^2}-{1\over k}$$ with $h<0$.  Following
\cite{FLSbig,LeSSii}, it is then straightforward to find the
weakly-interacting (large $u$) and strongly-interacting (small $u$)
series expansions of the current for general $k$. Defining the scaled
current ${\cal I}_k={I/gV}$ as in \cite{FS}, its UV expansion is
\begin{equation}
{\cal I}_k =
1+{\sqrt{\pi}\over 2}\sum_{n=1}^\infty (-1)^{n} 
{\Gamma(n(1+kh)+1)\Gamma(nh+1)\over n!\Gamma(nkh+1)\Gamma(nh+3/2)}
u^{2nh}
\label{currUV}
\end{equation}
Similarly, for the IR expansion, one finds
\begin{equation}
{\cal I}_k =
{\sqrt{\pi}\over 2}\sum_{n=1}^\infty (-1)^{n+1} 
{\Gamma(n(1+kh')+1)\Gamma(nh'+1)\over n!\Gamma(nkh'+1)\Gamma(nh'+3/2)}
u^{2nh'}.
\label{currIR}
\end{equation}
One can define a crossover temperature $T_B\propto
\lambda^{-1/h}$ analogous to the Kondo temperature $T_K$
\footnote{The exact relation between $T_B$ and the bare parameter
$\lambda$ can be found within the dimensional regularization scheme
usual in the TBA approach, see eg \cite{LeSSii}. 
One finds $u={V\over T_B'}$, with
$T_B'={N(-i)\over N(0)}e^{-\Delta} T_B$, $T_B=e^{\theta_B}$, and
$\theta_B$ the usual rapidity variable in the reflection matrix.}.

Remarkably, it turns out that this current can be written as the integral
\begin{equation}
{\cal I}_k(g,u)={i\over 4u}
\int_{{\cal C}_0} dx {(1+x^{h})^{k-1}\over \sqrt{x(1 + x^{h})^k - u^2}},
\label{claim}
\end{equation}
where the curve ${\cal C}_0$ starts at the origin, loops around the
branch point on the positive real axis, and goes back to the origin.
We derive the result (\ref{claim}) in the appendix. The current turns
out to obey the self-duality relation
\begin{equation}
{\cal I}_k(g,u)=1-{\cal I}_k\left({k^2\over g},u\right).
\label{duality}
\end{equation}
This follows directly from the expansions (\ref{currUV},\ref{currIR});
it can also easily be proven from the integral (\ref{claim}) by
changing variables $x\to x^{g/k}$ on the right-hand side and
integrating by parts. The physical origin of the duality is similar to
the $k=1$ case \cite{FS,LeS}. Integrability restricts the irrelevant
operators near the IR fixed point to be (within a dimensionally
regularized scheme) mutually commuting conserved quantities, either
neutral, or of the form
\begin{equation}
e^{i{k\over g}\phi}\chi_1
+e^{-i{k\over g}\phi}\chi_1^\dagger\label{dualparaf}.
\end{equation}
In particular, no harmonics of (\ref{dualparaf}) appear. It can also
be shown that the neutral quantities do not contribute to the DC
current, which is, in effect, completely determined by
(\ref{dualparaf}), hence giving rise to (\ref{duality}).

The representation (\ref{claim}) lets us easily find all the Lee-Yang singularities of the
current. Write ${\cal I}_k=\int {dx\over y}$. 
 Two or more roots of $y$ coalesce at values $u$ and $x$
where both $y=0$ and where $dy/dx =0$. The current will have a
singularity if the contour runs in between the coalescing roots. For
$k>1$, one such value is $u=0$, where $k$ roots coalesce at
$x^h=-1$. Since the contour ${\cal C}_0$ for the current is trivial
when $u$=0 (it starts at the origin and loops around the root at
$x$=0), the current is not singular. We will see in the next section
that the overscreened Kondo problem however has very interesting
behavior as a result of these $k$ roots coalescing.  All singularities
other than $u$=0 are at magnitude
$$|u_0|^2=(-h)^{k}\left(h+1\right)^{-k-{1\over h}};$$
(recall that $h$ is negative). For
real physical values of $u$, the contour never is
singular. The value of $|u_0|$ does give the radius of convergence
of the two perturbation expansions; the large-$u$ series converge for
$|u|>|u_0|$ and the small $u$ series for $|u|<|u_0|$.

\section{The multi-channel Kondo problem}

As is well known and discussed in depth in \cite{FLS}, the
single-channel Kondo problem and the Luttinger liquid with impurity
are deeply related. Not surprisingly, the multichannel problems 
are as well.

The Kondo model describes three-dimensional non-relativistic electrons
coupled to a single impurity spin. Considering the radial modes
reduces the problem to gapless electrons on the half-line coupled to a
quantum-mechanical spin ${\bf S}$ at the boundary. In the multichannel
Kondo problem, there are $k$ channels of electrons $\psi_{i\alpha}$
where $i=1\dots k$ and $\alpha=1,2$ is the spin index \cite{NB}.  One
can then form an $SU(2)_k$ ``spin'' current
$${\bf J}(x)=
 \sum_{i=1}^k \psi^{\dagger}_{i\alpha} 
{\bf \sigma}_{\alpha\beta} \psi_{i\beta}$$
using the Pauli matrices ${\bf \sigma}$. One can similarly form
$SU(k)_2$ ``flavor'' currents and a $U(1)$ ``charge'' current.
In conformal field theory language, there
are $2k$ Dirac fermions, which can be bosonized in terms of the
current algebras $SU(2)_k \times SU(k)_2 \times U(1)$ \cite{AL}. The
corresponding WZW theories have central charges $3k/(k+2),
2(k^2-1)/(k+2)$ and $1$ respectively, adding up to $2k$ as they
should.

Since only the spin current couples to the impurity, it is the only
one which we need here. Thus just like the multi-channel wire
considered in the previous section, the multi-channel Kondo model is
associated with $SU(2)_k$.  The impurity is represented by a
quantum-mechanical spin ${\bf S}$ in the spin-$S$ representation.  For
an impurity located at $x=0$, the fermions are coupled
antiferromagnetically via a term in the Hamiltonian
$$\delta H=\lambda {\bf J}(0)\cdot {\bf S}$$ for positive $\lambda$.
The coupling $\lambda$ is dimensionless since the current is of
dimension one, but there is a short-distance divergence in
perturbation theory in $\lambda$. Thus the interaction term is
marginally relevant, and a mass scale is present in the theory. In
particle-physics language, the Kondo model is asymptotically free and
undergoes dimensional transmutation.  This scale generated is usually
called the Kondo temperature $T_K$, and it is completely analogous to
$\Lambda_{QCD}$ in gauge theory.  In terms of the original parameter
$\lambda$ \cite{NB},
\begin{equation}
T_K \sim \lambda^{k/2}e^{-const/\lambda}.
\label{tklam}
\end{equation}
The renormalized theory parameter $T_K$ remains finite while the
bare parameter $\lambda\to 0$.

As $\lambda$ gets large (or more precisely, we study physics at energy
scales below $T_K$), the system crosses over to a strongly-coupled
phase.  At $T_K\to\infty$, there is another fixed point, where the the
electrons try to bind to the spin. Because of Pauli exclusion only a
single electron from each channel can bind to the impurity.  Thus the
problem naturally splits into three cases: overscreened ($k>2S$),
exactly screened ($k=2S$) and underscreened ($k<2S$). At this
strongly-coupled fixed point, the spin of the impurity is effectively
reduced to zero in the first two cases, while it is reduced to $S-k/2$
in the underscreened case. 

It is convenient to consider a more general model, the anisotropic
Kondo model, which allows for $SU(2)$-breaking interaction $\lambda_z
J_z S_z$. As detailed in \cite{FLS} for the single-channel case, this
remains integrable as long as the impurity spin is the appropriate
representation of the quantum group $SL(2,q)$. For $S=1/2$ this
distinction is irrelevant, since the representation is identical (the
Pauli matrices). As with the quantum wire, we parameterize the
anisotropy by the parameter $g$ with $0<g \le k$, where $g=k$
corresponds to the $SU(2)$-invariant isotropic point for the
$k$-channel problem. After a few simple transformations to gauge away
the $J_z S_z$ term and unfolding, the action for the spin degrees of
freedom reads much like in the previous section (\ref{fullint}), the
only difference being the impurity term, which is now of the form
\begin{equation}
\lambda\left( S^-e^{i\phi(0)}\chi_1(0)+S^+e^{-i\phi(0)}\chi^\dagger_1(0)\right)
\end{equation}
Here, $\phi$ is a spin boson (it was rather a charge boson in
(\ref{fullint})), and $S^\pm$ are raising and lowering $SL(2,q)$
operators in the appropriate spin $j$ representation. The deformation
parameter $q=e^{i\pi h}$, where $h={g\over k^2}-{1\over k}$ as with
the multichannel wire. The perturbing operators are not marginal as in
the isotropic case, but are relevant with dimension $d=1+h$. An
interesting point is the Toulouse point, where for $k=1$ and $g=1/2$,
or $k=2$ and $g=0$ \cite{EK}, the problem reduces to free fermions and
can be solved without recourse to the Bethe ansatz. For a review of
the current status of much of the theory and experiment of the
multi-channel Kondo model, see \cite{cox}.

The free energy in the multi-channel Kondo problem was derived using
the Bethe ansatz in \cite{TW,AD}. It is given in terms of the solution
of a set of an infinite number of non-linear integral equations. These
equations cannot be solved in closed form at arbitrary temperature,
but at zero temperature they reduce to a single linear integral
equation, which can be solved by the Wiener-Hopf technique.  The
physical quantity we will study is the magnetization $M_{k,S}$ of the
spin-$S$ impurity as a function of applied magnetic field $H$ (the
magnetic field couples to the conserved total charge $J_z+S_z$). At
zero temperature, the $M_{k,S}$ is a function only of the
dimensionless quantity $u$, where
$$u={g\Gamma(1/2h')\Gamma(k/2) \over 2\pi k^{k/2}\Gamma(k/2+1/2h')}
{H\over T_K}.$$
In the no-coupling limit, $M_{k,S}(u\to\infty)=S$, while in
the strong-coupling limit, $M_{k,S}(0)=0$ for the overscreened
case or exactly screened cases $k\ge 2S$ and $M_{k,S}(0)=(S-k/2)N/g$
for the underscreened cases $k< 2S$.

The entire magnetization for the isotropic case was derived in
\cite{TW} using the Bethe ansatz.  It is straightforward to generalize
this to all $g$; the result for the overscreened case and exactly
screened cases $k\ge 2S$ is
\begin{eqnarray}
\nonumber
M_{k,S}\left(u\right)&&=\frac{i}{4\pi^{3/2}}
\int_{-\infty}^\infty
\frac{d\omega}{\omega-i\epsilon} e^{2i\omega\ln(u)}
\frac{\sinh(2S\pi\omega)}{\sinh(k\pi\omega)}\\
&&\qquad\qquad\times
\frac{\Gamma(i\omega)\Gamma({1\over 2}-i\omega)\Gamma(1-i\omega/h)}
{\Gamma(ik\omega)\Gamma(1-i\omega g/kh)}
\label{mag}
\end{eqnarray}
where $\epsilon$ is positive and tending to zero.
While this expression is somewhat unwieldy, it is easy to find
complete perturbative expansions from it by completing the contour in
the upper half-plane for $u$ large, and in the lower half-plane
for $u$ small. The poles in the upper half-plane are at
$\omega=-inh$ for $n\ge 0$ an integer (recall that $h<0$). Thus for $u$
large enough (the precise limit will be given below), the
magnetization for spin $1/2$  is
\begin{equation}
M_{k,1/2}={1\over 2\sqrt{\pi}}\sum_{n=0}^\infty {(-1)^n\over n!}
{\Gamma({1\over 2}-nh)\Gamma(1-knh)\over
\Gamma(1-nh) \Gamma(1-nkh-n)}u^{2nh}
\label{magUV}
\end{equation}
In the isotropic case $h\to 0$, this expansion breaks down
because the exponent goes to zero while $u\to 0$ as $h^k$ for fixed
$H/T_K$. In this case, the appropriate expansion involves $\ln (H/T_K)$, as is
clear from (\ref{tklam}).

The main result of this section is that the magnetization in the
$k$-channel Kondo problem can be expressed simply in terms of
a hyperelliptic curve:
\begin{equation}
M_{k,S}(u)= \frac{iu}{4\pi}\int_{{\cal C}_S} \frac{dx}{xy}
\label{bigone}
\end{equation}
where
\begin{equation}
y^2 = (-1)^{2S}x(1-x^h)^k +  u^2
\label{ykondo}
\end{equation}
and the contour ${\cal C}_{S}$ starts at infinity and goes around the
``first'' 2S branch points. The hyperelliptic curve (\ref{ykondo})
differs by that in the previous section only by minus signs. This
seemingly innocuous change is responsible for the interesting
non-fermi-liquid behavior in the overscreened Kondo problem.

The derivation of this curve for $S=1/2$ starting from the series
expansion (\ref{magUV}) is similar to that of the appendix,
so we omit it here. The contour ${\cal C}_{1/2}$ starts at infinity
(as opposed to the origin for ${\cal C}_0$), loops around the branch point
on the real axis for real positive $u$, and returns to infinity.  The
derivation of the contours for higher impurity spins is identical to
the derivation for $k=1$ in \cite{Paul}.  Since $S=1/2$ is the most
interesting physical case, and the contours for higher spin were
discussed in detail in \cite{Paul}, we give only a brief explanation
here. The higher-spin magnetizations follows from the ``fusion''
relation valid at large $u$:
\begin{equation}
M_{k,S}(iu) + M_{k,S}(-iu)= M_{k,S-1/2}(u) + M_{k,S+1/2}(u)
\label{fusion}
\end{equation}
where the argument $iu$ is meant as the continuous deformation of
$u\to iu$ at fixed large $|u|$. For the overscreened case $k>2S$, this
relation is actually valid at all $u$.  As one can see from the form
of the expansion (\ref{magUV}), $M(e^{2\pi i}u)\ne M(u)$.  The
``first'' $2S$ branch points are then those obtained by rotating the
original branch point on the real axis by $e^{2\pi ij}$ for
$j=-(S-1/2)\dots (S+1/2)$. These points are on different sheets
because of the branch cut along the negative $x$-axis due to the $x^h$
in (\ref{ykondo}). For example, for $S=1$ and large $u$, on the
original sheet there are branch points just above and below the
negative real axis near $x\approx -u^2$. Thus when $u$ is large, the contour
${\cal C}_1$ starts at infinity, loops around the upper branch point,
returns to infinity and then loops around the other branch point. This
can be deformed to a closed contour surrounding the two branch points
on the original sheet. By analytic continuation, this contour is valid
for any value of $u$, not just $u$ large where the formula (\ref{fusion})
applies.

The behavior near the strong-coupling fixed point is easy to find
using the curve. Because of the $(1-x^h)^k$ in (\ref{ykondo}), for $u$
small there are $k$ branch points near $x=1$. As opposed to the current
in the previous section, the contour here involves these points.

For the
exactly-screened case $k=2S$, the contour surrounds just these $k$ branch
points. Expanding the integrand in (\ref{bigone}) in powers of $u^2$ and
using the beta-function identity (\ref{betaid}) yields the expansion
\begin{equation}
M_{k,k/2}=\sum_{n=0}^\infty a_n u^{2n+1}
\label{exactIR}
\end{equation}
where
$$a_n={1\over h\sqrt{\pi}}
{(-1)^n\Gamma(-(n+{1\over 2})/h)\Gamma(n+1/2)\over
n! \Gamma(1-(n+{1\over 2})g/kh) \Gamma(k(n+{1\over 2}))}$$
good for small $u$. This of course agrees with the result obtained
from completing the contour in (\ref{mag}) in the lower
half-plane. Physically, this expansion means that the irrelevant
operators near the IR fixed point are all scalar, of even dimensions:
powers of the stress energy tensor and the like, exactly as in the
$k=1$ case.

The under-screened case is very similar to the
single-channel case discussed in \cite{Paul}.  The irrelevant operators
near the IR fixed point then involve not only the foregoing scalar operators, 
but also, like in the tunneling problem, a single charged operator
\begin{equation}
S^- e^{ik\phi/g}\chi_1+ S^+ e^{-ik\phi/g}
\chi_1^\dagger\label{dualkondo},
\end{equation}
where $S$ now is a $SL(2,q)$ spin in the $j-{k\over 2}$ representation. 
The dimension of this operator is $d=1+h'={k-1\over k}+{1\over g}$. 
The form (\ref{dualkondo}) establishes a 
duality between the strong and weak coupling regimes, similar to 
what happens for $k=1$ \cite{FW,Paul}. This duality is only ``partial'',
because, in contrast to the current in the previous section, 
the magnetization, when expanded in the IR, depends not only
on the term (\ref{dualkondo}), but also 
on the scalar irrelevant operators.

\begin{figure}
\epsfxsize=3.5in
\epsffile{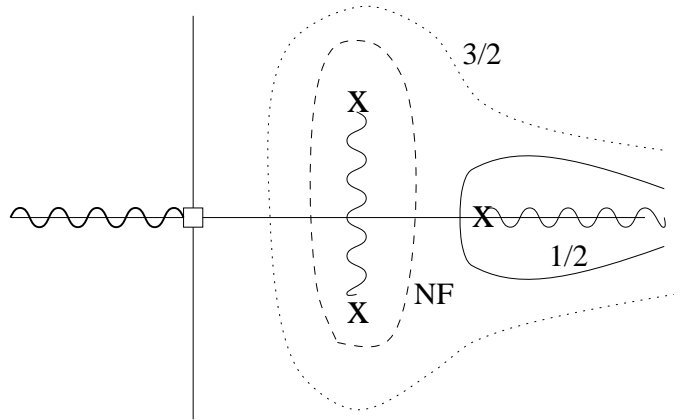}
\caption{The contours for spin $1/2$ (overscreened) and spin $3/2$ (exactly
screened) impurities for the three-channel Kondo problem. The three
square-root branch points illustrated all approach $x=1$ as $u\to 0$.}
\end{figure}

The overscreened case is the most interesting because of the
non-fermi-liquid behavior even in the isotropic limit $h\to 0$.  The
curve is singular at $u=0$ because $k$ roots are coalescing at
$x=1$. The magnetization is singular as well because the contour runs
in between these coalescing roots (for the under-screened and exactly
screened cases, the contour surrounds all of these and so is not
singular). For the example $k=3$, this is illustrated in figure 1; the
roots pictured are those which meet at $x=1$ when $u=0$. This singular
behavior when the curve goes in between these roots results in a
non-fermi critical exponent. We illustrate this first for odd $k$. The
contour for $S=1/2$ can be written as the sum of two contours, the
exactly-screened contour ${\cal C}_{k/2}$ plus a contour ${\cal
C}_{\rm NF}$ surrounding the other $k-1$ branch points near $x=1$. The
exactly-screened contour of course yields the expansion
(\ref{exactIR}) with its Fermi-liquid exponent. To find the
appropriate expansion for the contour ${\cal C}_{\rm NF}$, we change
variables in (\ref{bigone}) by $r=(1-x^h)u^{-2/k}$, so for odd $k$
\begin{eqnarray}
\nonumber
M_{k,1/2}&=&(-1)^{(k-1)/2}M_{k,k/2}\\
\nonumber
&&+{i\over 4\pi h}\int_{{\cal C}_{\rm NF}} 
\frac{dr}{1-u^{2/k}r}
\frac{u^{2/k}}{\sqrt{1-r^k(1-u^{2/k}r)^{1/h}}}.
\end{eqnarray}
This can be expanded in powers of $u^{2/k}$ when $u$ is small, giving
the appropriate non-fermi critical exponent \cite{TW,AD}. To find the
coefficients of this expansion, one divides the contour ${\cal C}_{\rm
NF}$ into $k-1$ contours each starting and ending at the origin $r=0$,
and again utilizes a computation similar to the appendix. The result
is
\begin{eqnarray}
\nonumber
&&M_{k,1/2}=(-1)^{(k-1)/2}M_{k,k/2}\\
\nonumber
&&\ +{1\over 2\pi^{1/2}}
\sum_{n=1}^\infty
\frac{\Gamma({1\over 2}-n/k)\Gamma(1-n/kh)}{n!\Gamma(1-n/k)\Gamma(1-n/kh-n)}
u^{2n/k}
\label{overIR}
\end{eqnarray}
of course in agreement with the residue expansion of (\ref{mag}).
This expansion is still valid in the isotropic limit $h=0$.

These results are interpreted physically as follows. In addition to
the scalar quantities, there is another irrelevant operator
controlling the approach to the IR fixed point in the overscreened
case, and replacing (\ref{dualkondo}), of the form
\begin{equation}
\epsilon_1\partial\phi\label{irrelevope}
\end{equation}
where $\epsilon_1$ is the energy field of the $Z_k$ parafermionic
theory, of dimension $d={2\over k+2}$ \cite{FZ}.  There is a slight
subtlety concerning the computation of the magnetization near the IR
fixed point in the overscreened case, since the impurity spin has
disappeared right at the fixed point \cite{AL}.  At zero temperature,
if we call $\lambda_d$ the coupling of (\ref{irrelevope}), it turns
out that the magnetization goes as $(H^d\lambda_d)^{1/(1-d)}$
\cite{AL}, so indeed the magnetization goes as $u^{2/k}$.

For $k$ even, although the curve is still simple, the small-coupling
expansion is cumbersome because there are an even number
of roots coalescing. For example, for $k=2$ and $u$ small, the leading
term is
\begin{equation}
M_{2,1/2}\approx
\frac{iu}{4\pi h}
\int_{1+u}^{const} {dx\over x}{1\over\sqrt{u^2-(x-1)^2}}
\propto u\ln u.
\end{equation}
The fusion relation (\ref{fusion}) lets us see right away that the log
terms must be related to the expansion (\ref{exactIR}) for the exactly
screened problem. In general, for $k$ even, the expansion is of the form
\begin{equation}
M_{k,1/2}=(-1)^{(k-1)/2}\sum_{n=1}^\infty (-1)^n{a_n\over \pi} u^{2n+1}\ln u 
+ b_n u^{2n/k}
\end{equation}
where the $a_n$ are given above, and the $b_n$ are quite complicated
(involving a sum of digamma functions).
The easiest way to find the $b_n$ is in
fact to go back to the original Bethe ansatz expression (\ref{mag}).

We finally discuss the isotropic limit $g=k$ ($h\to 0$), which is
particularly intriguing. In this limit $u\to 0$ as $h^{k/2}$ for fixed
$H/T_K$.  Denoting ${\cal M}_{k,S}(H/T_K)=\lim_{g\to 1} M_{k,S}(u)$,
its integral form is
$$
{\cal M}_{k,S}
=\frac{i}{4\pi}\int_{{\cal C}_S} \frac{dx}{x}
\frac{H/T_K}{\sqrt{(-1)^{2S}2\pi x(\ln x)^k + (H/T_K)^2}}.
$$
Thus it is obvious why the weak-coupling perturbation expansion around
$T_K=0$ involves logarithmic terms. To formulate a large $H/T_K$ perturbation
theory, we define the parameter
\begin{equation}
\ln(H/T_K)=\frac{1}{z}-\frac{k}{2}\ln (z/4\pi).
\label{zdef}
\end{equation}
By a change of variables, the magnetization can be written for $g\to 1$ as
\begin{equation}
M_{k,S} =\frac{i}{4\pi}\int_{{\cal C}_{S}} \frac{dx}{x}
\frac{1}{\left[(-1)^{2S}x((z/2)\ln x -1)^k+1\right]^{1/2}}
\label{magz}
\end{equation}
As with the single-channel case, this can be expanded in powers of $z$:
\begin{equation}
{\cal M}_{k,S}(z)=\sum_{n=0}^\infty{\cal A}_n z^n
\label{expz}
\end{equation}
For spin $S=1/2$ this expansion is asymptotic. It has zero radius of
convergence because as $x\to\infty$ the $x(z\ln x)^k$ term will
eventually dominate the integral no matter how small $z$ is.
In contrast, observe that the multi-channel quantum wire discussed in
the previous section is very simple at $h\to 0$; the integral for the
current can easily be done explicitly. The perturbing operator
(\ref{delH}) for the wire is exactly marginal at this value of $h$, so
the model remains conformally invariant even with the perturbation.
Therefore no scale is introduced, and the problem can be
solved using the techniques of boundary conformal field theory.

\bigskip
This research was supported by NSF grant DMR-9802813 (P.F.), the NYI program 
(NSF-PHY-9357207)
and DOE grant DE-FG03-84ER40168 (H.S.)

  \appendix
\section{The derivation of the curve}

We consider the expression
$$f(\lambda)=
{i\over 2}\int_{{\cal C}_0} dx~{\left(1+\lambda x^h\right)^{k-1}\over \left[
x\left(1+\lambda x^h\right)^k-1\right]^{1/2}},$$
where the $x$ integral is along the usual contour starting at the
origin, looping around the branch point on the real axis when
$\lambda$ is real, and ending at the origin. We represent $f(\lambda)$
as a double integral
$${i\over 2}\int\int dxdz ~{z^{k-1}\over\left(xz^k-1\right)^{1/2}}
~\delta\left[z-\left(1+\lambda x^h\right)\right].$$
We can now expand the square root by setting $xz^k-1=x-1-x(1-z^k)$, leading to
\begin{eqnarray}
\nonumber
\int\int dxdz~ z^{k-1}&&\sum_{n=0}^\infty (-1)^n {\Gamma(n+1/2)\over
\Gamma(1/2)\Gamma(n+1)}~{x^n\left(1-z^k\right)^n\over (1-x)^{n+1/2}}\\
\nonumber 
&&\qquad\times\delta\left[z-\left(1+\lambda x^h\right)\right].
\end{eqnarray}
We now represent the delta function as a third integral
$$\delta\left[z-\left(1+\lambda x^h\right)\right]=
\int dt ~ e^{2i\pi t\left(z-1-\lambda x^h\right)},$$
and expand the term $e^{-2i\pi t\lambda x^h}$ in power series. Each
$x$ integral is done by using the contour-integral representation of
the beta function
\begin{equation}
\frac{\Gamma(a)}{\Gamma(a+b)\Gamma(1-b)} = \frac{i}{2\pi}
\int_{{\cal C}_{0}} dx\ x^{a-1} (x-1)^{b-1},
\label{betaid}
\end{equation}
yielding
\begin{eqnarray}
\nonumber
\int dzdt\ && e^{2i\pi t(z-1)} z^{k-1}\sum_{n=0}^\infty 
\sum_{p=0}^\infty
\left(1-z^k\right)^n\\
\nonumber
&&\times{\Gamma(n+ph+1)\over \Gamma(1/2)\Gamma(n+1)\Gamma(ph+3/2)} 
{(-2i\pi t\lambda)^p\over p!}.
\end{eqnarray}
To remove the $t^p$ piece, we integrate by parts $p$
times in $z$.  The $t$ integral then yields $\delta(z-1)$, which then
lets us do the $z$ integral, yielding
$$
{1\over \sqrt{\pi}}\sum_{p=0}^\infty
\lambda^p~{\Gamma(ph+1)\over \Gamma(ph+3/2)p!}~ C_p,
$$
where 
$$C_p=\sum_{n=0}^\infty {\Gamma(n+ph+1)\over \Gamma(ph+1) n!}
\left. {d^p\over dz^p}~ z^{k-1}\left(1-z^k\right)^n\right|_{z=1}.$$
The sum over $n$ actually truncates above at $n=p$ because of the
$z$=1 limit, and below at $n=(p-k+1)/k$ because of the
derivatives. However, it is more convenient to leave the bounds $0$
and $\infty$, because the sum in $C_p$ can then be done:
\begin{eqnarray}
\nonumber
C_p&=&\left. {d^p\over dz^p}\left(z^{k-1} \sum_{n=0}^\infty
{\Gamma(n+ph+1)\over \Gamma(ph+1) n!}   (1-z^k)^n\right) \right|_{z=1}\\
\nonumber
&=&\left.{d^p\over dz^p}\left(z^{k-1} z^{-k(ph+1)}\right)\right|_{z=1}\\
\nonumber
&=&(-1)^p{\Gamma(kph+p+1)\over\Gamma(kph+1)}.
\end{eqnarray}
It follows that
\begin{equation}
f(\lambda)={1\over\sqrt{\pi}}
\sum_{p=0}^\infty (-1)^p~ \lambda^{2p}~ {\Gamma(ph+1)\Gamma(kph+p+1)
\over p!\Gamma(ph+3/2)\Gamma(kph+1)}.
\end{equation}
Of course, the above manipulations are true only for values of
$\lambda$ where the series converges. Using a straightforward change
of variables one finds the final form
\begin{equation}
{\cal I}={i\over 4u}\int_{C_0} dx~
{\left(1+x^h\right)^{k-1}\over \sqrt{
x\left(1+ x^h\right)^k-u^2}}.
\end{equation}

\end{document}